%% file: Dv2lett-cern.tex
\documentclass[ALICE,manyauthors]{cernphprep}
\usepackage[english]{babel}
\usepackage{graphicx}
\usepackage{verbatim}
\usepackage{amsfonts}
\usepackage{makeidx}
\usepackage{color}
\usepackage{amsmath}
\usepackage{lineno}
\usepackage{epstopdf}
\usepackage{hyperref}
\usepackage{cite}
\newcommand{\pp}{\rm pp}

\newcommand{\sqrtsNN}{\sqrt{s_{\rm \scriptscriptstyle NN}}}

\newcommand{\GeV}{\mathrm{GeV}}
\newcommand{\TeV}{\mathrm{TeV}}

\newcommand{\mev}{\mathrm{MeV}}
\newcommand{\gev}{\mathrm{GeV}}
\newcommand{\tev}{\mathrm{TeV}}

\newcommand{\PbPb}{\mbox{Pb--Pb}}

\newcommand{\RAA}{R_{\rm AA}}

\newcommand{\pt}{p_{\rm T}}
\newcommand{\vtwo}{v_{\rm 2}}
\newcommand{\DeltaPhi}{\Delta\varphi}

\newcommand{\DtoKpi}{{\rm D^0\to K^-\pi^+}}
\newcommand{\DtoKpipi}{{\rm D^+\to K^-\pi^+\pi^+}}
\newcommand{\DstartoDpi}{{\rm D^{*+}\to D^0\pi^+}}
\newcommand{\Dzero}{{\rm D^0}}
\newcommand{\Dstar}{{\rm D^{*+}}}
\newcommand{\Dplus}{{\rm D^+}}

\newcommand{\dEdx}{{\rm d}E/{\rm d}x}
\newcommand{\dNdphi}{{\rm d}N/{\rm d}\varphi}

\newcommand{\jpsi}{{\rm J/}\psi}
\begin{document}

\begin{titlepage}

\PHnumber{2013-085}                 
\PHdate{May 10, 2013}              

\title{D meson elliptic flow in non-central $\PbPb$ collisions at 
$\sqrtsNN~=~2.76~\TeV$}
\Collaboration{ALICE Collaboration%
         \thanks{See Appendix~\ref{app:collab} for the list of collaboration 
                      members}}
\ShortAuthor{ALICE Collaboration}
\ShortTitle{D meson elliptic flow in non-central $\PbPb$ collisions at 
$\sqrtsNN~=~2.76~\TeV$}

\begin{abstract}
Azimuthally anisotropic distributions of $\Dzero$, $\Dplus$ and $\Dstar$ 
mesons were studied in the central rapidity 
region ($|y|<0.8$) in Pb--Pb collisions at a centre-of-mass energy 
$\sqrtsNN=2.76~\tev$ per nucleon--nucleon collision,
with the ALICE detector at the LHC.
The second Fourier coefficient $v_2$ (commonly denoted  
elliptic flow) was measured in the centrality class 30--50\% as a 
function of the D meson transverse momentum $\pt$, in the range 2--16~$\gev/c$.
The measured $v_2$ of D mesons is comparable in magnitude to that of 
light-flavour hadrons. 
It is positive in the range $2<\pt<6$~$\gev/c$ with 
5.7\,$\sigma$ significance, based on the combination of statistical and 
systematic uncertainties.

\end{abstract}

\end{titlepage}
\setcounter{page}{2}

\input{intro.tex}

\input{dataandreco.tex}

\input{methods.tex}

\input{syst.tex}

\input{results.tex}

\newenvironment{acknowledgement}{\relax}{\relax}
\begin{acknowledgement}
\section*{Acknowledgements}
\input{acknowledgements_march2013.tex}
\end{acknowledgement}

\vspace{1cm}

\input{references.tex}
\newpage
\appendix
\section{The ALICE Collaboration}
\label{app:collab}
\input{authorlist-2013-05-02-cernpreprint.tex}  

\end{document}

%% file: intro.tex
Heavy-ion collisions at ultra-relativistic energies 
are aimed at exploring the structure of nuclear matter at 
extremely high temperatures and energy densities.
Under these conditions, according to Quantum-Chromodynamics (QCD) calculations
on the lattice, the confinement of quarks and gluons inside hadrons is 
no longer effective and a phase transition to a Quark-Gluon Plasma (QGP) 
occurs~\cite{karsch}.

The measurement of anisotropy in the azimuthal distribution of particle
momenta provides insight into the properties of the QGP medium.
Anisotropy in particle momenta originates from the initial 
anisotropy in the spatial distribution of the nucleons 
participating in the collision.
The anisotropy of produced particles 
is characterized by the Fourier
coefficients $v_n=\langle\cos[n(\varphi-\Psi_n)]\rangle$,
where $\varphi$ is the azimuthal angle of 
the particle, and $\Psi_n$ is the azimuthal angle of the initial state 
symmetry plane for the $n$-th harmonic.
For non-central collisions the overlap region of the colliding nuclei 
has a lenticular shape and the anisotropy 
is dominated by the second coefficient $v_2$, commonly denoted 
elliptic flow~\cite{ALICEv2, ALICEhighharm}.

The $v_2$ values measured at RHIC and LHC can be described by the combination of
two mechanisms~\cite{STARv2,STARv2id,PHENIXhighptv2,hydroVisc,ALICEv2,CMSv2,ATLASv2,CMShighptv2,ALICEhighptv2,whdg2011}.
The first one, dominant at low ($\pt<3~\gev/c$) and intermediate 
(3--6~$\gev/c$) transverse momentum, is the build-up of a collective expansion 
through interactions among the medium constituents.
Elliptic flow develops mainly in the early stages of this collective 
expansion, when the spatial anisotropy is 
large~\cite{Ollitrault,hydro,Heinz}.
The second mechanism 
is the path-length dependence of in-medium parton energy loss, due to
medium-induced gluon radiation and elastic collisions. 
This is predicted to give rise to a positive $v_2$ for 
hadrons up to large $\pt$~\cite{highptv2a,highptv2b}.

The measurement of the elliptic flow of charmed hadrons provides further
insight into the transport properties of the medium.
In contrast to light quarks and gluons that can be produced or annihilated 
during the entire evolution of the medium, heavy quarks are produced 
predominantly in initial hard scattering processes and their annihilation rate 
is expected to be small~\cite{PBM}.
Hence, the final state heavy-flavour hadrons at all transverse 
momenta originate from heavy quarks that experienced all stages of the 
system evolution.
At low $\pt$, charmed hadron $v_2$ offers a unique opportunity to test whether 
also quarks with large mass ($m_{\rm c}\approx 1.5~\gev/c^2$) 
par\-ti\-ci\-pa\-te in the collective expansion dynamics and possibly 
thermalize in the medium~\cite{Batsouli,GrecoKoRapp}.
Because of their large mass, charm quarks are expected to have a longer
relaxation time, i.e.\ time scale for approaching equilibrium with the 
medium, with respect to light quarks~\cite{Teaney}.
At low and intermediate $\pt$, the D meson elliptic flow is expected to
be sensitive to the heavy-quark hadronization mechanism.
In case of substantial interactions with the medium, a significant 
fraction of low- and intermediate-momentum heavy quarks could hadronize via 
recombination with other quarks from the bulk of thermalized 
partons~\cite{Molnar,GrecoKoLevai}, thus enhancing the $v_2$ of D mesons 
with respect to that of charm quarks~\cite{GrecoKoRapp}.
In this context, the measurement of D meson $v_2$ is also relevant for the 
interpretation of the results on $\jpsi$ anisotropy~\cite{jpsiv2}, because 
$\jpsi$'s from ${\rm c\overline{c}}$ (re)combination would inherit 
the anisotropy of their constituent 
quarks~\cite{LiuXuZhuang,ZhaoEmerickRapp}. 
At high $\pt$, the D meson $v_2$ can constrain the path-length dependence of
parton energy loss, complementing the measurement of the 
nuclear modification factor $\RAA$~\cite{RAA}, 
defined as the ratio of the yield in nucleus--nucleus to that observed
in $\pp$ collisions scaled by the number of binary nucleon--nucleon 
collisions.
A large suppression of the inclusive D meson yield ($\RAA \approx 0.25$)
is observed in central Pb--Pb collisions at $\sqrtsNN=2.76~\tev$ for 
$\pt>5~\gev/c$~\cite{aliceDRAA}.

Theoretical models of heavy-quark interactions with the medium constituents 
predict, for semi-central collisions at the LHC, a large $v_2$ (0.1--0.2) 
for $\rm D$ mesons at $\pt\approx 2$--3~GeV/$c$ 
and a decrease to about 0.05 at high 
$\pt$~\cite{Gossiaux,BAMPS,Rapp,Beraudo,Lang}.
The elliptic flow of electrons from heavy-flavour decays 
was measured to be as large as 0.13 in Au--Au collisions at 
$\sqrtsNN=200~\gev$~\cite{phenixHFEv2,phenixHFERAAv2}.

In this Letter we present the measurement of $v_2$ for
$\Dzero$, $\Dplus$, and $\Dstar$ mesons and their anti-particles
reconstructed from their hadronic decays at mid-rapidity ($|y|<0.8$) in 
non-central Pb--Pb collisions at $\sqrtsNN=2.76~\tev$.

%% file: dataandreco.tex
The measurement was carried out with the ALICE detector at the 
LHC~\cite{aliceJINST}.
Particle reconstruction and identification were based on the 
detectors of the 
central barrel, located inside a solenoid magnet, which generates 
a 0.5~T field parallel to the beam direction.

The detectors used for the reconstruction of the trajectories of candidate
D meson decay particles are the Inner Tracking System (ITS), 
composed of six cylindrical layers of silicon detectors~\cite{ITSalign},
and the Time Projection Chamber (TPC)~\cite{TPCpaper}.
The reconstructed particles are identified on the basis of their specific 
energy deposition $\dEdx$ in the TPC gas and of their time-of-flight
from the interaction point to the Time Of Flight (TOF) detector.
The ITS, TPC and TOF detectors provide full azimuthal coverage in the 
pseudorapidity interval $|\eta|<0.9$.

The analysis was performed on a sample of $\PbPb$ collisions 
collected in 2011 with an interaction trigger that required coincident signals 
in both scintillator arrays of the VZERO detector, covering the full 
azimuth in the regions $-3.7 < \eta < -1.7$ and $2.8 < \eta < 5.1$.
Events were further selected off\-line to remove background from 
beam-gas interactions, using the time information provided by the 
VZERO and the neutron Zero-Degree Calorimeters (ZDC).
Only events with a vertex reconstructed within $\pm$10~cm from the centre
of the detector along the beam line were considered in the analysis.
Collisions were classified according to their centrality, 
determined from the VZERO summed amplitudes and defined in terms of 
percentiles of the total hadronic $\PbPb$ cross section~\cite{ALICEcentr}. 

The D meson $v_2$ was measured for
events in the centrality range 30--50\%, where the initial geometrical 
anisotropy and the medium density are large.
In this range, the trigger and event selection are fully efficient for 
hadronic interactions.
The number of selected events in the 30--50\% centrality class 
was $9.5\times10^6$,
corresponding to an integrated luminosity of $(6.2\pm0.2)~\mu{\rm b}^{-1}$.

The decays
$\DtoKpi$, $\DtoKpipi$ and $\DstartoDpi$, and their charge conjugates, were
reconstructed as described in~\cite{aliceDpp7,aliceDRAA}. 
$\Dzero$ and $\Dplus$ candidates were formed using pairs and triplets of 
tracks with $|\eta|<0.8$, $\pt>0.4~\gev/c$, at least 70 associated space 
points in the TPC, and at least two hits in the ITS, of which 
at least one should be in either of the two innermost layers.
$\Dstar$ candidates were formed by combining $\Dzero$ candidates
with tracks with $|\eta|<0.8$, $\pt>0.1~\gev/c$, and at least three associated 
hits in the ITS.
The selection of tracks with $|\eta|<0.8$ limits the D meson acceptance
in rapidity, which, depending on $\pt$, varies from $|y|<0.7$ for
$\pt = 2~\GeV/c$ to $|y|<0.8$ for $\pt > 5~\GeV/c$.

D meson candidates were selected with the same stra\-te\-gy as used 
in~\cite{aliceDRAA}, in order to increase the 
statistical significance of the signal with respect to the large
background of all possible track combinations.
The selection of the decay topology was based on the displacement
of the decay tracks from the interaction vertex, the separation 
between the secondary and primary vertices, and 
the pointing of the reconstructed D meson momentum to the primary vertex.
The pion and kaon identification in the TPC and TOF detectors was utilized 
by applying cuts in units of resolution
(at $\pm3\,\sigma$) around the expected mean values of $\dEdx$ and 
time-of-flight.

%% file: methods.tex
The measurement of $v_2$ 
was performed by correlating the candidate D meson azi\-mu\-thal angle, 
$\varphi_{\rm D}$, 
with the angle $\psi_2$ of the so-called event plane~\cite{PoskVol}, which is
an estimator of the direction $\Psi_2$ of the second-order initial-state 
symmetry plane.
The event plane angle $\psi_2$ was determined from the second harmonic of the 
azimuthal distribution of the detected charged particles:
\mbox{$\psi_2\ = \frac{1}{2} \tan^{-1}\left(Q_{2,y}/Q_{2,x}\right)$},
where $Q_{2,x}$ and $Q_{2,y}$ are the transverse components of the second order
flow vector, $\vec{Q}_2$,
defined event-by-event from the azimuthal angles $\varphi_i$ of a sample of 
$N$ tracks,
\mbox{$\vec{Q}_2=\left( \sum_{i=1}^{N} w_i \cos 2\varphi_i~,\sum_{i=1}^{N} w_i \sin 2\varphi_i\right)$}.
The weights $w_i$ correct for 
non-uniformities in the acceptance and efficiency of the 
detector, and optimize the event-plane resolution~\cite{PoskVol}.
They are defined as the product of the track $\pt$ and the inverse of the 
probability of reconstructing a particle with azimuthal angle $\varphi_i$.
The tracks used to compute $\vec{Q}_2$ were required to have at least 
50 associated space points in the TPC, $0 < \eta < 0.8$,
$\pt>150~\mev/c$,
and distance of closest approach to the primary vertex smaller than 3.2~cm 
along the beam direction and 2.4~cm in the transverse plane.
To avoid auto-correlations between the D mesons and the event plane, 
the angle $\psi_2$ was recalculated for each candidate after 
subtracting from the $\vec{Q}_2$ vector the 
contribution from the tracks used to form that particular candidate.
A correlation of D mesons with the tracks used to determine the 
event plane could also originate from other sources, commonly denoted  
non-flow, which are not related to the correlation with the initial 
geometry symmetry plane, such as higher-mass particle decays or jets.
Their effect was estimated to be small with respect to the other uncertainties
by repeating the analysis using the event plane determined in a different 
$\eta$ region with the VZERO detector.

\begin{figure}[!hbt]
\begin{center}
\includegraphics[width=0.4\textwidth]{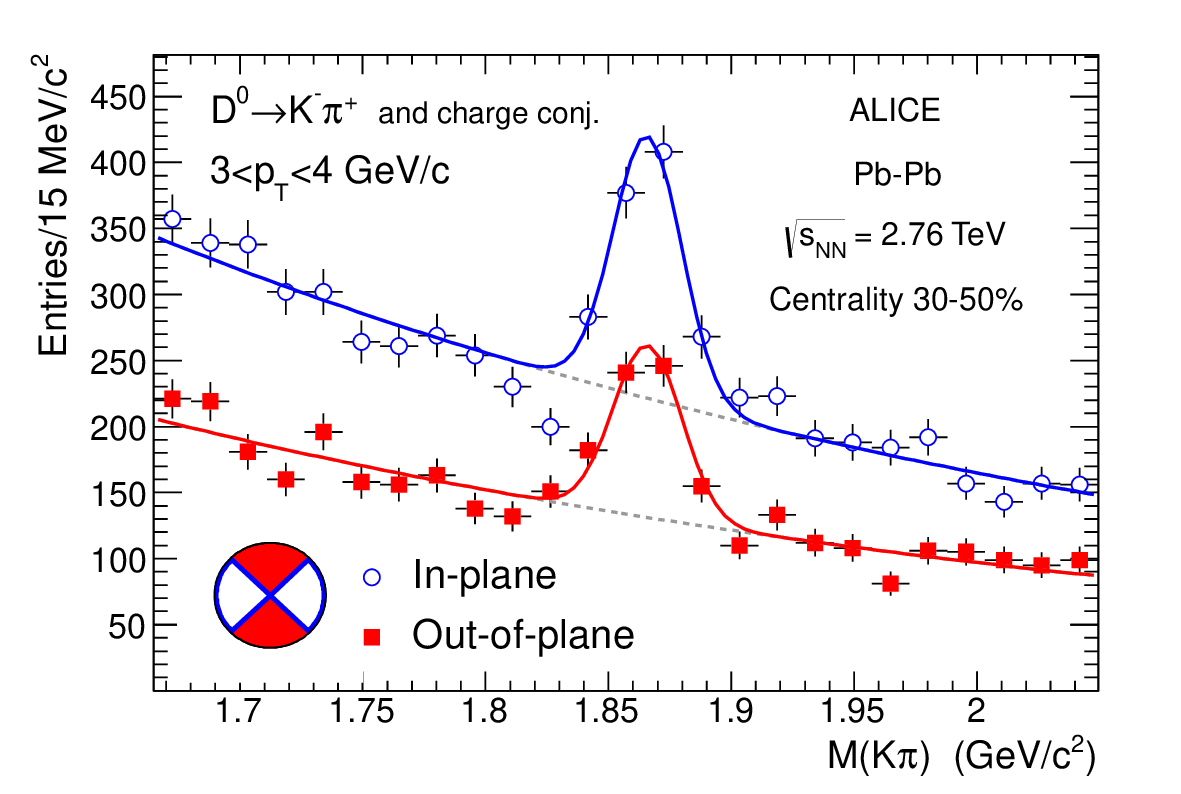}
\caption{Invariant mass distributions for $\Dzero$ candidates and their
charge conjugates with $3<\pt<4~\gev/c$ for $9.5\times10^6$ $\PbPb$ 
collisions in the 30--50\% centrality class. 
The distributions are shown separately for the in-plane (open symbols) and 
out-of-plane (closed symbols) intervals of azimuthal angle.
The curves show the fit functions as described in the text.}
\label{fig:invmass3} 
\end{center}
\end{figure}

D meson candidates were classified in two groups according to 
their azimuthal angle relative to the event plane 
($\Delta\varphi=\varphi_{\rm D}-\psi_2$): in-plane
($\left]-\frac{\pi}{4},\frac{\pi}{4}\right]$ and
 $\left]\frac{3\pi}{4},\frac{5\pi}{4}\right]$)
and out-of-plane
($\left]-\frac{\pi}{4},\frac{3\pi}{4}\right]$ and
 $\left]\frac{5\pi}{4},\frac{7\pi}{4}\right]$).

The raw signal yields were extracted in each $\Delta\varphi$ and $\pt$ interval
by means of a fit to the candidate invariant mass distributions 
(mass difference $M({\rm K} \pi\pi) -M({\rm K} \pi)$ for $\Dstar$).
The fitting function was the sum of a Gaussian function to describe the 
signal and an exponential (for $\Dzero$ and 
$\Dplus$) or a power-law (for $\Dstar$) function for the background.
An example fit is shown in Fig.~\ref{fig:invmass3} for
$\Dzero$ candidates.
For each meson and in each $\pt$ interval, the mean and the width of the 
Gaussian were fixed to those obtained from a fit to the invariant mass 
distribution integrated over $\Delta\varphi$, whose signal peak has larger
statistical significance.
The raw yields in the two $\Delta\varphi$ intervals,
$N_{\rm in\mbox{-}plane}$ and $N_{\rm out\mbox{-}of\mbox{-}plane}$ were obtained as the 
integrals over the corresponding Gaussian signal functions.
$v_2$ was computed as:
\begin{equation} \label{eq:v2}
v_2=\frac{1}{R_2}\frac{\pi}{4}\frac{N_{\rm in\mbox{-}plane}-N_{\rm out\mbox{-}of\mbox{-}plane}}{N_{\rm in\mbox{-}plane}+N_{\rm out\mbox{-}of\mbox{-}plane}}\,.
\end{equation}
The factor $\pi/4$ results from the integration of the
second term, $2\,v_2\cos(2\Delta\varphi)$, of the $\dNdphi$ distribution
in the considered $\Delta\varphi$ intervals and the factor $1/R_2$ is the 
correction for the finite resolution in the estimation of the symmetry plane 
$\Psi_2$ via the event plane $\psi_2$~\cite{PoskVol}.
$R_2$ was determined from the correlation between the
event plane angles calculated from tracks reconstructed in the two sides 
of the TPC, namely $-0.8<\eta<0$ and $0<\eta<0.8$.
The resulting value is $R_2=0.8059\pm0.0001$(stat)$\pm0.024$(syst).

The measured D meson yield has a contribution from feed-down from B meson 
decays, which amounts to about 10--20\%~\cite{aliceDRAA,aliceDpp7}, 
depending on the selection cuts and $\pt$.
Indeed, the B feed-down contribution is enhanced by the 
selection criteria that are more efficient for feed-down D mesons, because
their decay vertices are more displaced from the primary vertex.
Thus, the measured $v_2$ is a combination of those
of promptly produced and of feed-down D mesons. 
Considering that the elliptic flow is additive, 
the value for promptly produced D mesons, $v_2^{\rm prompt}$, can be obtained 
from the measured $v_2^{\rm all}$ as:
\begin{equation}
\label{eq:bfeed}
v_2^{\rm prompt}=\frac{v_2^{\rm all}}{f_{\rm prompt}} -\frac{1-f_{\rm prompt}}{f_{\rm prompt}}v_2^{\rm feed\mbox{-}down}~,
\end{equation}
where $f_{\rm prompt}$ is the fraction of promptly produced D mesons in the 
measured raw yield and $v_2^{\rm feed\mbox{-}down}$ is the elliptic flow of D mesons
from B decays, which depends on the dynamics of beauty quarks in the medium.
These two quantities have not been measured.
However, as it can be seen in Eq.~(\ref{eq:bfeed}), 
$v_2^{\rm all}$ coincides with $v_2^{\rm prompt}$, independent of 
$f_{\rm prompt}$, if $v_2^{\rm feed\mbox{-}down}=v_2^{\rm prompt}$.
The assumption $v_2^{\rm feed\mbox{-}down}=v_2^{\rm prompt}$ was used to
compute the central value of the results for the prompt D meson elliptic flow.
The systematic uncertainty related to this assumption is discussed below.

%% file: syst.tex
The contributions to the systematic uncertainty on the measured $v_2$ originate from:
\begin{itemize}
\item determination of D meson yields and their 
anisotropy relative to the event plane (e.g.\ 10--30\% in $4<\pt<6~\gev/c$
depending on the meson species); 
\item non-flow effects and centrality dependence in the event plane resolution ($3\%$); 
\item B feed-down contribution (typically $^{+45}_{-0}\%$).
\end{itemize}
 
\begin{figure*}[!t]
\begin{center}
\includegraphics[width=.9\textwidth]{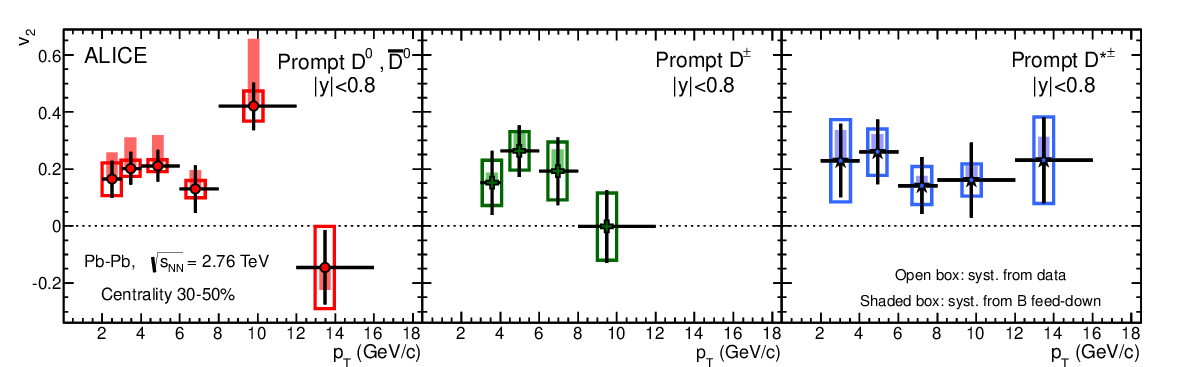}
\caption{$\vtwo$ as a function of $\pt$ for prompt $\Dzero$, $\Dplus$ and 
$\Dstar$ mesons for $\PbPb$ collisions in the centrality range 30--50\%.
The central value was obtained with the assumption 
$v_2^{\rm feed\mbox{-}down}=v_2^{\rm prompt}$.
Vertical error bars represent the statistical uncertainty, empty boxes the 
systematic uncertainty due to the D meson anisotropy measurement and the 
event-plane resolution, and shaded boxes show the uncertainty from the 
contribution of D mesons from B feed-down.}
\label{fig:v2withhadrons3pad} 
\end{center}
\end{figure*}

The first contribution was estimated from the maximum deviation from the 
central $v_2$ value
obtained by repeating the yield extraction in each $\pt$ and $\DeltaPhi$ 
interval when varying the fit configuration: different fit functions were 
used for the background; the Gaussian width and mean were left as free 
parameters in the fit; the yield 
was defined by counting the histogram entries in the invariant 
mass region of the signal, after subtracting the background contribution
estimated from a fit to the side bands.

The $v_2$ result obtained with 
Eq.~(\ref{eq:v2}) was cross checked by using an independent technique based on 
fits to the measured $\vtwo$ of 
candidates as a function of their invariant mass, $M$~\cite{Borghini}.
Here $\vtwo(M)$ was obtained 
with methods based on two-particle correlations, namely the scalar 
product~\cite{SP} and the $Q$-cumulants~\cite{QC}.

It was checked that the results were stable against variations of the 
cuts applied for the selection of D meson candidates, and that the 
reconstruction and selection efficiencies from Monte Carlo simulations 
were compatible for the in-plane and out-of-plane D mesons.

The uncertainty on the correction factor $R_2$ for the event plane resolution
has two contributions. 
The first one is due to the centrality dependence of $R_2$.
The average $R_2$ in the 30-50\% centrality interval was computed assuming 
that the D meson yield is uniformly distributed as a function of centrality.
A systematic uncertainty of 2\% was assigned by comparing this value with an
alternative estimation of the average where the $R_2$ values in narrow 
centrality intervals were weighted with the D meson yields measured in the 
same intervals.
The second contribution to the $R_2$ uncertainty arises from the presence of
non-flow correlations between the two sub-events used to compute the 
resolution. 
The systematic uncertainty was estimated to be of 2.3\% on the basis of
the difference to the $R_2$ value obtained
using three sub-events with a wider pseudorapidity gap, namely TPC tracks and 
the signals in the two VZERO detectors.

\begin{figure}[!b]
\begin{center}
\includegraphics[width=0.5\textwidth]{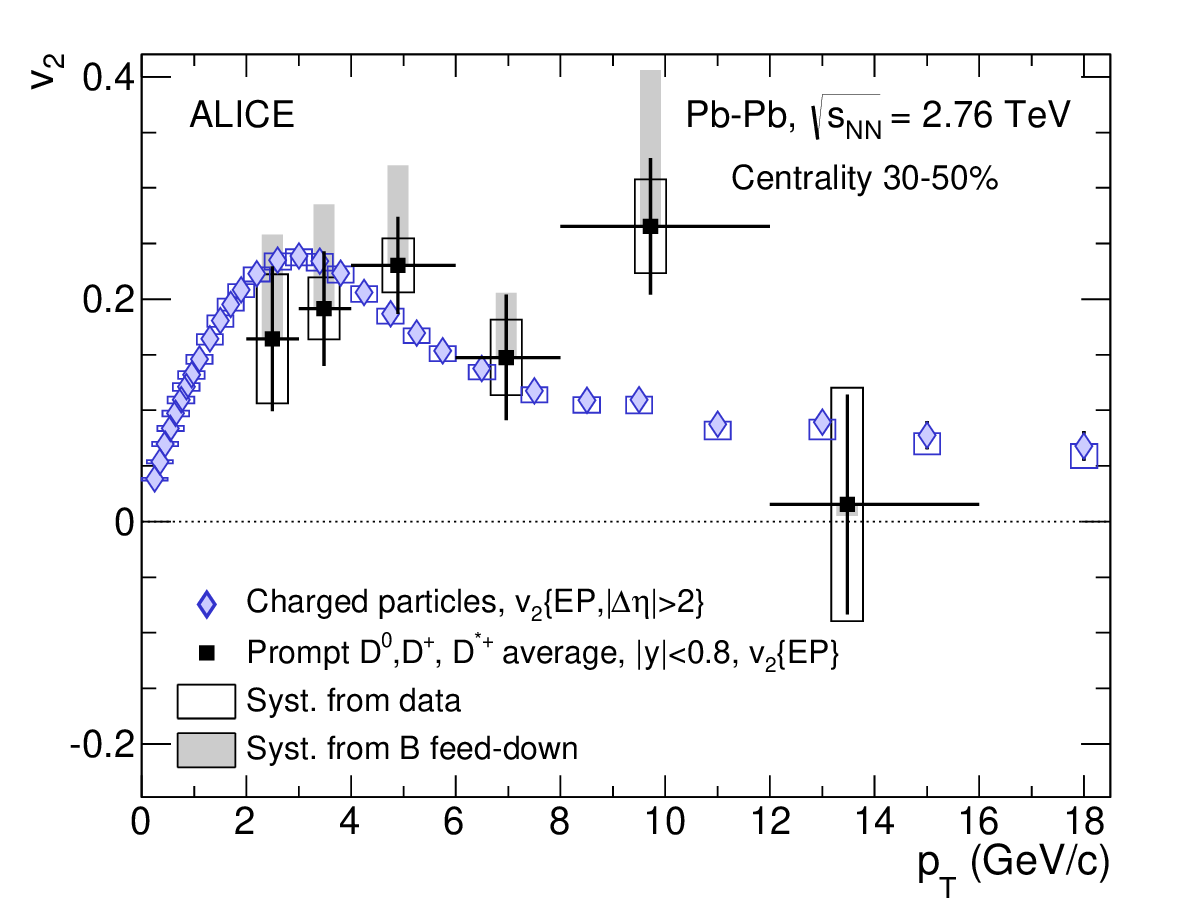}
\caption{Average of $\Dzero$, $\Dplus$ and $\Dstar$ $v_2$ as a function of 
$\pt$, compared to charged-particle $v_2$~\cite{ALICEhighptv2} measured with the 
event plane (EP) method.
The symbols are positioned horizontally at the average $\pt$ of the three
D meson species.}
\label{fig:v2withhadrons} 
\end{center}
\end{figure}

The systematic uncertainty related to the
contribution of  D mesons from B decays was assigned by varying the assumption 
on the elliptic flow of feed-down D mesons in Eq.~(\ref{eq:bfeed}) in the 
range $0 \le v_2^{\rm feed\mbox{-}down} \le v_2^{\rm prompt}$, which 
includes all model predictions~\cite{Gossiaux, BAMPS, Rapp, Lang}.
The maximum variation corresponds to the case $v_2^{\rm feed\mbox{-}down}=0$, 
which gives $v_2^{\rm prompt} =v_2^{\rm all}/f_{\rm prompt}$.
Hence, the magnitude of the systematic uncertainty due to B feed-down 
is inversely proportional to $f_{\rm prompt}$.
We estimated $f_{\rm prompt}$ as described in~\cite{aliceDRAA} using: 
(i) FONLL~\cite{FONLL} predictions for prompt D and B 
mesons; (ii) B$\rightarrow$D+$X$ decay kinematics from EvtGen~\cite{EvtGen}; 
(iii) reconstruction and selection efficiencies for 
prompt and feed-down D mesons from simulations; and (iv) a hypothesis on the 
nuclear modification factor of the feed-down D mesons, 
$\RAA^{\rm feed\mbox{-}down}$.
The latter factor accounts for the medium-induced modification of 
the $\pt$ distribution of B mesons.
Its contribution was determined by varying the ratio  
$\RAA^{\rm feed\mbox{-}down}/\RAA^{\rm prompt}$ in the range 1--3,
motivated by the lower value of $\RAA$ of prompt D mesons measured
by ALICE~\cite{aliceDRAA} with respect to preliminary results from the CMS
experiment on the $\RAA$ of $\rm J/\psi$ from B decays~\cite{CMSsecjpsi}.
The B feed-down uncertainty was defined by the lower limit
of the resulting $f_{\rm prompt}$ range, which depends on the D
meson species and $\pt$.
A typical value for this lower limit is 0.68, corresponding
to a relative uncertainty on $v_2^{\rm prompt}$ of $^{+45}_{-0}\%$.

%% file: results.tex
Figure~\ref{fig:v2withhadrons3pad} shows the measured $v_2$ as a function of 
$\pt$ for $\Dzero$, $\Dplus$ and $\Dstar$ mesons
in the 30--50\% centrality class. 
The symbols are positioned horizontally at the average $\pt$ of 
reconstructed D mesons, determined as described in~\cite{aliceDpp7}. 
The elliptic flow of the three D meson species is consistent within 
uncertainties.
An average $v_2$, and transverse momentum, of $\Dzero$, $\Dplus$ and $\Dstar$ was 
computed using the statistical uncertainties as weights. 
The systematic uncertainties were propagated 
through the averaging procedure,
treating the contributions from the event-plane re\-so\-lu\-tion
and the B feed-down correction as fully correlated among the 
three D meson species.
The resulting D meson $v_2$ is shown in Fig.~\ref{fig:v2withhadrons}.
It is comparable in magnitude to that of charged 
particles, dominated by light-flavour hadrons~\cite{ALICEhighptv2}.
The average of the measured D meson $v_2$ values in the interval 
$2<\pt<6~\gev/c$ is
$0.204\pm 0.030$~(stat)~$\pm 0.020$~(syst)~$_{-0}^{+0.092}$~(B feed-down),
which is larger than zero with $5.7\,\sigma$ significance.
This result indicates that the interactions with the medium constituents 
transfer to charm quarks information on the azimuthal anisotropy of the system,
suggesting that low momentum charm quarks take part in the 
collective motion of the system.
A positive $v_2$ is also observed for $\pt>6~\gev/c$, which most likely 
originates from the path-length dependence of the partonic energy loss, 
although the large uncertainties do not allow for a firm conclusion.

In summary, we have presented the first measurement of the D meson elliptic 
flow coefficient $v_2$ 
for semi-central Pb--Pb collisions at $\sqrtsNN=2.76~\tev$.
A positive elliptic flow, with $5.7\,\sigma$ significance in $2<\pt<6~\gev/c$, 
is observed.
This $v_2$ measurement, together with the observed large suppression of 
D mesons in central collisions~\cite{aliceDRAA}, provides a stringent 
constraint to theoretical models describing the interaction of heavy quarks 
with the medium.

%% file: acknowledgements_march2013.tex
The ALICE collaboration would like to thank all its engineers and technicians for their invaluable contributions to the construction of the experiment and the CERN accelerator teams for the outstanding performance of the LHC complex.
\\
The ALICE Collaboration would like to thank M.~Cacciari for 
providing the pQCD predictions used for the feed-down correction.
\\
The ALICE collaboration acknowledges the following funding agencies for their support in building and
running the ALICE detector:
 \\
State Committee of Science,  World Federation of Scientists (WFS)
and Swiss Fonds Kidagan, Armenia,
 \\
Conselho Nacional de Desenvolvimento Cient\'{\i}fico e Tecnol\'{o}gico (CNPq), Financiadora de Estudos e Projetos (FINEP),
Funda\c{c}\~{a}o de Amparo \`{a} Pesquisa do Estado de S\~{a}o Paulo (FAPESP);
 \\
National Natural Science Foundation of China (NSFC), the Chinese Ministry of Education (CMOE)
and the Ministry of Science and Technology of China (MSTC);
 \\
Ministry of Education and Youth of the Czech Republic;
 \\
Danish Natural Science Research Council, the Carlsberg Foundation and the Danish National Research Foundation;
 \\
The European Research Council under the European Community's Seventh Framework Programme;
 \\
Helsinki Institute of Physics and the Academy of Finland;
 \\
French CNRS-IN2P3, the `Region Pays de Loire', `Region Alsace', `Region Auvergne' and CEA, France;
 \\
German BMBF and the Helmholtz Association;
\\
General Secretariat for Research and Technology, Ministry of
Development, Greece;
\\
Hungarian OTKA and National Office for Research and Technology (NKTH);
 \\
Department of Atomic Energy and Department of Science and Technology of the Government of India;
 \\
Istituto Nazionale di Fisica Nucleare (INFN) and Centro Fermi -
Museo Storico della Fisica e Centro Studi e Ricerche "Enrico
Fermi", Italy;
 \\
MEXT Grant-in-Aid for Specially Promoted Research, Ja\-pan;
 \\
Joint Institute for Nuclear Research, Dubna;
 \\
National Research Foundation of Korea (NRF);
 \\
CONACYT, DGAPA, M\'{e}xico, ALFA-EC and the EPLANET Program
(European Particle Physics Latin American Network)
 \\
Stichting voor Fundamenteel Onderzoek der Materie (FOM) and the Nederlandse Organisatie voor Wetenschappelijk Onderzoek (NWO), Netherlands;
 \\
Research Council of Norway (NFR);
 \\
Polish Ministry of Science and Higher Education;
 \\
National Authority for Scientific Research - NASR (Autoritatea Na\c{t}ional\u{a} pentru Cercetare \c{S}tiin\c{t}ific\u{a} - ANCS);
 \\
Ministry of Education and Science of Russian Federation, Russian
Academy of Sciences, Russian Federal Agency of Atomic Energy,
Russian Federal Agency for Science and Innovations and The Russian
Foundation for Basic Research;
 \\
Ministry of Education of Slovakia;
 \\
Department of Science and Technology, South Africa;
 \\
CIEMAT, EELA, Ministerio de Econom\'{i}a y Competitividad (MINECO) of Spain, Xunta de Galicia (Conseller\'{\i}a de Educaci\'{o}n),
CEA\-DEN, Cubaenerg\'{\i}a, Cuba, and IAEA (International Atomic Energy Agency);
 \\
Swedish Research Council (VR) and Knut $\&$ Alice Wallenberg
Foundation (KAW);
 \\
Ukraine Ministry of Education and Science;
 \\
United Kingdom Science and Technology Facilities Council (STFC);
 \\
The United States Department of Energy, the United States National
Science Foundation, the State of Texas, and the State of Ohio.

%% file: authorlist-2013-05-02-cernpreprint.tex
\begingroup
\small
\begin{flushleft}
B.~Abelev\Irefn{org1234}\And
J.~Adam\Irefn{org1274}\And
D.~Adamov\'{a}\Irefn{org1283}\And
A.M.~Adare\Irefn{org1260}\And
M.M.~Aggarwal\Irefn{org1157}\And
G.~Aglieri~Rinella\Irefn{org1192}\And
M.~Agnello\Irefn{org1313}\textsuperscript{,}\Irefn{org1017688}\And
A.G.~Agocs\Irefn{org1143}\And
A.~Agostinelli\Irefn{org1132}\And
Z.~Ahammed\Irefn{org1225}
% ...
\renewcommand\labelenumi{\textsuperscript{\theenumi}~}
\section*{Affiliation notes}
\renewcommand\theenumi{\roman{enumi}}
\begin{Authlist}
\item \Adef{M.V.Lomonosov Moscow State University, D.V.Skobeltsyn Institute of Nuclear Physics, Moscow, Russia}Also at: M.V.Lomonosov Moscow State University, D.V.Skobeltsyn Institute of Nuclear Physics, Moscow, Russia
\item \Adef{University of Belgrade, Faculty of Physics and "Vinvca" Institute of Nuclear Sciences, Belgrade, Serbia}Also at: University of Belgrade, Faculty of Physics and "Vinvca" Institute of Nuclear Sciences, Belgrade, Serbia
\item \Adef{Deceased}Also at: Deceased
\item \Adef{Institute of Theoretical Physics, University of Wroclaw, Wroclaw, Poland}Also at: Institute of Theoretical Physics, University of Wroclaw, Wroclaw, Poland
\end{Authlist}
\section*{Collaboration Institutes}
\renewcommand\theenumi{\arabic{enumi}~}
\begin{Authlist}
\item \Idef{org36632}Academy of Scientific Research and Technology (ASRT), Cairo, Egypt
\item \Idef{org1332}A. I. Alikhanyan National Science Laboratory (Yerevan Physics Institute) Foundation, Yerevan, Armenia
\item \Idef{org1279}Benem\'{e}rita Universidad Aut\'{o}noma de Puebla, Puebla, Mexico
\item \Idef{org1220}Bogolyubov Institute for Theoretical Physics, Kiev, Ukraine
\item \Idef{org20959}Bose Institute, Department of Physics and Centre for Astroparticle Physics and Space Science (CAPSS), Kolkata, India
\item \Idef{org1262}Budker Institute for Nuclear Physics, Novosibirsk, Russia
\item \Idef{org1292}California Polytechnic State University, San Luis Obispo, California, United States
\item \Idef{org1329}Central China Normal University, Wuhan, China
\item \Idef{org14939}Centre de Calcul de l'IN2P3, Villeurbanne, France
\item \Idef{org1197}Centro de Aplicaciones Tecnol\'{o}gicas y Desarrollo Nuclear (CEADEN), Havana, Cuba
\item \Idef{org1242}Centro de Investigaciones Energ\'{e}ticas Medioambientales y Tecnol\'{o}gicas (CIEMAT), Madrid, Spain
\item \Idef{org1244}Centro de Investigaci\'{o}n y de Estudios Avanzados (CINVESTAV), Mexico City and M\'{e}rida, Mexico
\item \Idef{org1335}Centro Fermi - Museo Storico della Fisica e Centro Studi e Ricerche ``Enrico Fermi'', Rome, Italy
\item \Idef{org17347}Chicago State University, Chicago, United States
\item \Idef{org1288}Commissariat \`{a} l'Energie Atomique, IRFU, Saclay, France
\item \Idef{org15782}COMSATS Institute of Information Technology (CIIT), Islamabad, Pakistan
\item \Idef{org1294}Departamento de F\'{\i}sica de Part\'{\i}culas and IGFAE, Universidad de Santiago de Compostela, Santiago de Compostela, Spain
\item \Idef{org1106}Department of Physics Aligarh Muslim University, Aligarh, India
\item \Idef{org1121}Department of Physics and Technology, University of Bergen, Bergen, Norway
\item \Idef{org1162}Department of Physics, Ohio State University, Columbus, Ohio, United States
\item \Idef{org1300}Department of Physics, Sejong University, Seoul, South Korea
\item \Idef{org1268}Department of Physics, University of Oslo, Oslo, Norway
\item \Idef{org1315}Dipartimento di Fisica dell'Universit\`{a} and Sezione INFN, Trieste, Italy
\item \Idef{org1145}Dipartimento di Fisica dell'Universit\`{a} and Sezione INFN, Cagliari, Italy
\item \Idef{org1312}Dipartimento di Fisica dell'Universit\`{a} and Sezione INFN, Turin, Italy
\item \Idef{org1285}Dipartimento di Fisica dell'Universit\`{a} `La Sapienza' and Sezione INFN, Rome, Italy
\item \Idef{org1154}Dipartimento di Fisica e Astronomia dell'Universit\`{a} and Sezione INFN, Catania, Italy
\item \Idef{org1132}Dipartimento di Fisica e Astronomia dell'Universit\`{a} and Sezione INFN, Bologna, Italy
\item \Idef{org1270}Dipartimento di Fisica e Astronomia dell'Universit\`{a} and Sezione INFN, Padova, Italy
\item \Idef{org1290}Dipartimento di Fisica `E.R.~Caianiello' dell'Universit\`{a} and Gruppo Collegato INFN, Salerno, Italy
\item \Idef{org1103}Dipartimento di Scienze e Innovazione Tecnologica dell'Universit\`{a} del Piemonte Orientale and Gruppo Collegato INFN, Alessandria, Italy
\item \Idef{org1114}Dipartimento Interateneo di Fisica `M.~Merlin' and Sezione INFN, Bari, Italy
\item \Idef{org1237}Division of Experimental High Energy Physics, University of Lund, Lund, Sweden
\item \Idef{org1192}European Organization for Nuclear Research (CERN), Geneva, Switzerland
\item \Idef{org1227}Fachhochschule K\"{o}ln, K\"{o}ln, Germany
\item \Idef{org1122}Faculty of Engineering, Bergen University College, Bergen, Norway
\item \Idef{org1136}Faculty of Mathematics, Physics and Informatics, Comenius University, Bratislava, Slovakia
\item \Idef{org1274}Faculty of Nuclear Sciences and Physical Engineering, Czech Technical University in Prague, Prague, Czech Republic
\item \Idef{org1229}Faculty of Science, P.J.~\v{S}af\'{a}rik University, Ko\v{s}ice, Slovakia
\item \Idef{org1184}Frankfurt Institute for Advanced Studies, Johann Wolfgang Goethe-Universit\"{a}t Frankfurt, Frankfurt, Germany
\item \Idef{org1215}Gangneung-Wonju National University, Gangneung, South Korea
\item \Idef{org20958}Gauhati University, Department of Physics, Guwahati, India
\item \Idef{org1212}Helsinki Institute of Physics (HIP) and University of Jyv\"{a}skyl\"{a}, Jyv\"{a}skyl\"{a}, Finland
\item \Idef{org1203}Hiroshima University, Hiroshima, Japan
\item \Idef{org1254}Indian Institute of Technology Bombay (IIT), Mumbai, India
\item \Idef{org36378}Indian Institute of Technology Indore, Indore, India (IITI)
\item \Idef{org1266}Institut de Physique Nucl\'{e}aire d'Orsay (IPNO), Universit\'{e} Paris-Sud, CNRS-IN2P3, Orsay, France
\item \Idef{org1277}Institute for High Energy Physics, Protvino, Russia
\item \Idef{org1249}Institute for Nuclear Research, Academy of Sciences, Moscow, Russia
\item \Idef{org1320}Nikhef, National Institute for Subatomic Physics and Institute for Subatomic Physics of Utrecht University, Utrecht, Netherlands
\item \Idef{org1250}Institute for Theoretical and Experimental Physics, Moscow, Russia
\item \Idef{org1230}Institute of Experimental Physics, Slovak Academy of Sciences, Ko\v{s}ice, Slovakia
\item \Idef{org1127}Institute of Physics, Bhubaneswar, India
\item \Idef{org1275}Institute of Physics, Academy of Sciences of the Czech Republic, Prague, Czech Republic
\item \Idef{org1139}Institute of Space Sciences (ISS), Bucharest, Romania
\item \Idef{org27399}Institut f\"{u}r Informatik, Johann Wolfgang Goethe-Universit\"{a}t Frankfurt, Frankfurt, Germany
\item \Idef{org1185}Institut f\"{u}r Kernphysik, Johann Wolfgang Goethe-Universit\"{a}t Frankfurt, Frankfurt, Germany
\item \Idef{org1177}Institut f\"{u}r Kernphysik, Technische Universit\"{a}t Darmstadt, Darmstadt, Germany
\item \Idef{org1256}Institut f\"{u}r Kernphysik, Westf\"{a}lische Wilhelms-Universit\"{a}t M\"{u}nster, M\"{u}nster, Germany
\item \Idef{org1246}Instituto de Ciencias Nucleares, Universidad Nacional Aut\'{o}noma de M\'{e}xico, Mexico City, Mexico
\item \Idef{org1247}Instituto de F\'{\i}sica, Universidad Nacional Aut\'{o}noma de M\'{e}xico, Mexico City, Mexico
\item \Idef{org1308}Institut Pluridisciplinaire Hubert Curien (IPHC), Universit\'{e} de Strasbourg, CNRS-IN2P3, Strasbourg, France
\item \Idef{org1182}Joint Institute for Nuclear Research (JINR), Dubna, Russia
\item \Idef{org1199}Kirchhoff-Institut f\"{u}r Physik, Ruprecht-Karls-Universit\"{a}t Heidelberg, Heidelberg, Germany
\item \Idef{org20954}Korea Institute of Science and Technology Information, Daejeon, South Korea
\item \Idef{org1017642}KTO Karatay University, Konya, Turkey
\item \Idef{org1160}Laboratoire de Physique Corpusculaire (LPC), Clermont Universit\'{e}, Universit\'{e} Blaise Pascal, CNRS--IN2P3, Clermont-Ferrand, France
\item \Idef{org1194}Laboratoire de Physique Subatomique et de Cosmologie (LPSC), Universit\'{e} Joseph Fourier, CNRS-IN2P3, Institut Polytechnique de Grenoble, Grenoble, France
\item \Idef{org1187}Laboratori Nazionali di Frascati, INFN, Frascati, Italy
\item \Idef{org1232}Laboratori Nazionali di Legnaro, INFN, Legnaro, Italy
\item \Idef{org1125}Lawrence Berkeley National Laboratory, Berkeley, California, United States
\item \Idef{org1234}Lawrence Livermore National Laboratory, Livermore, California, United States
\item \Idef{org1251}Moscow Engineering Physics Institute, Moscow, Russia
\item \Idef{org1322}National Centre for Nuclear Studies, Warsaw, Poland
\item \Idef{org1140}National Institute for Physics and Nuclear Engineering, Bucharest, Romania
\item \Idef{org1017626}National Institute of Science Education and Research, Bhubaneswar, India
\item \Idef{org1165}Niels Bohr Institute, University of Copenhagen, Copenhagen, Denmark
\item \Idef{org1109}Nikhef, National Institute for Subatomic Physics, Amsterdam, Netherlands
\item \Idef{org1283}Nuclear Physics Institute, Academy of Sciences of the Czech Republic, \v{R}e\v{z} u Prahy, Czech Republic
\item \Idef{org1264}Oak Ridge National Laboratory, Oak Ridge, Tennessee, United States
\item \Idef{org1189}Petersburg Nuclear Physics Institute, Gatchina, Russia
\item \Idef{org1170}Physics Department, Creighton University, Omaha, Nebraska, United States
\item \Idef{org1157}Physics Department, Panjab University, Chandigarh, India
\item \Idef{org1112}Physics Department, University of Athens, Athens, Greece
\item \Idef{org1152}Physics Department, University of Cape Town and  iThemba LABS, National Research Foundation, Somerset West, South Africa
\item \Idef{org1209}Physics Department, University of Jammu, Jammu, India
\item \Idef{org1207}Physics Department, University of Rajasthan, Jaipur, India
\item \Idef{org1200}Physikalisches Institut, Ruprecht-Karls-Universit\"{a}t Heidelberg, Heidelberg, Germany
\item \Idef{org1017688}Politecnico di Torino, Turin, Italy
\item \Idef{org1325}Purdue University, West Lafayette, Indiana, United States
\item \Idef{org1281}Pusan National University, Pusan, South Korea
\item \Idef{org1176}Research Division and ExtreMe Matter Institute EMMI, GSI Helmholtzzentrum f\"ur Schwerionenforschung, Darmstadt, Germany
\item \Idef{org1334}Rudjer Bo\v{s}kovi\'{c} Institute, Zagreb, Croatia
\item \Idef{org1298}Russian Federal Nuclear Center (VNIIEF), Sarov, Russia
\item \Idef{org1252}Russian Research Centre Kurchatov Institute, Moscow, Russia
\item \Idef{org1224}Saha Institute of Nuclear Physics, Kolkata, India
\item \Idef{org1130}School of Physics and Astronomy, University of Birmingham, Birmingham, United Kingdom
\item \Idef{org1338}Secci\'{o}n F\'{\i}sica, Departamento de Ciencias, Pontificia Universidad Cat\'{o}lica del Per\'{u}, Lima, Peru
\item \Idef{org1155}Sezione INFN, Catania, Italy
\item \Idef{org1313}Sezione INFN, Turin, Italy
\item \Idef{org1271}Sezione INFN, Padova, Italy
\item \Idef{org1133}Sezione INFN, Bologna, Italy
\item \Idef{org1146}Sezione INFN, Cagliari, Italy
\item \Idef{org1316}Sezione INFN, Trieste, Italy
\item \Idef{org1115}Sezione INFN, Bari, Italy
\item \Idef{org1286}Sezione INFN, Rome, Italy
\item \Idef{org36377}Nuclear Physics Group, STFC Daresbury Laboratory, Daresbury, United Kingdom
\item \Idef{org1258}SUBATECH, Ecole des Mines de Nantes, Universit\'{e} de Nantes, CNRS-IN2P3, Nantes, France
\item \Idef{org35706}Suranaree University of Technology, Nakhon Ratchasima, Thailand
\item \Idef{org1304}Technical University of Split FESB, Split, Croatia
\item \Idef{org1017659}Technische Universit\"{a}t M\"{u}nchen, Munich, Germany
\item \Idef{org1168}The Henryk Niewodniczanski Institute of Nuclear Physics, Polish Academy of Sciences, Cracow, Poland
\item \Idef{org17361}The University of Texas at Austin, Physics Department, Austin, TX, United States
\item \Idef{org1173}Universidad Aut\'{o}noma de Sinaloa, Culiac\'{a}n, Mexico
\item \Idef{org1296}Universidade de S\~{a}o Paulo (USP), S\~{a}o Paulo, Brazil
\item \Idef{org1149}Universidade Estadual de Campinas (UNICAMP), Campinas, Brazil
\item \Idef{org1239}Universit\'{e} de Lyon, Universit\'{e} Lyon 1, CNRS/IN2P3, IPN-Lyon, Villeurbanne, France
\item \Idef{org1205}University of Houston, Houston, Texas, United States
\item \Idef{org20371}University of Technology and Austrian Academy of Sciences, Vienna, Austria
\item \Idef{org1222}University of Tennessee, Knoxville, Tennessee, United States
\item \Idef{org1310}University of Tokyo, Tokyo, Japan
\item \Idef{org1318}University of Tsukuba, Tsukuba, Japan
\item \Idef{org21360}Eberhard Karls Universit\"{a}t T\"{u}bingen, T\"{u}bingen, Germany
\item \Idef{org1225}Variable Energy Cyclotron Centre, Kolkata, India
\item \Idef{org1017687}Vestfold University College, Tonsberg, Norway
\item \Idef{org1306}V.~Fock Institute for Physics, St. Petersburg State University, St. Petersburg, Russia
\item \Idef{org1323}Warsaw University of Technology, Warsaw, Poland
\item \Idef{org1179}Wayne State University, Detroit, Michigan, United States
\item \Idef{org1143}Wigner Research Centre for Physics, Hungarian Academy of Sciences, Budapest, Hungary
\item \Idef{org1260}Yale University, New Haven, Connecticut, United States
\item \Idef{org15649}Yildiz Technical University, Istanbul, Turkey
\item \Idef{org1301}Yonsei University, Seoul, South Korea
\item \Idef{org1327}Zentrum f\"{u}r Technologietransfer und Telekommunikation (ZTT), Fachhochschule Worms, Worms, Germany
\end{Authlist}
\endgroup